\documentclass[fleqn,usenatbib]{mnras}

\usepackage{newtxtext,newtxmath}

\usepackage[T1]{fontenc}
\usepackage{ae,aecompl}

\usepackage{graphicx}	
\usepackage{amsmath}	
\usepackage{amssymb}	

\title[Thick discs in AGN]{Extreme AGN variability: evidence of
  magnetically elevated accretion?}

\author[Dexter \& Begelman]{Jason Dexter$^{1}$\thanks{E-mail: jdexter@mpe.mpg.de} and Mitchell C. Begelman$^{2,3}$\\
$^{1}$Max-Planck-Institut f\"ur Extraterrestrische Physik, Giessenbachstr. 1, 85748 Garching, Germany\\
$^{2}$JILA, University of Colorado and NIST, 440 UCB, Boulder, CO 80309-0440, USA\\
$^{3}$Department of Astrophysical and Planetary Sciences, 391 UCB, University of Colorado, Boulder, CO 80309-0391, USA\\
}

\date{Accepted XXX. Received YYY; in original form ZZZ}

\pubyear{2018}

\begin{document}
\label{firstpage}
\pagerange{\pageref{firstpage}--\pageref{lastpage}}
\maketitle

\begin{abstract}
Rapid, large amplitude variability at optical to X-ray wavelengths is now seen in
an increasing number of Seyfert galaxies and luminous quasars. The
variations imply a global change in accretion power,
but are too rapid to be communicated by inflow through a standard thin
accretion disc. Such discs are long known to have difficulty
explaining the observed optical/UV emission from active galactic
nuclei. Here we show that alternative models developed to explain
these observations have larger scale heights and shorter inflow
times. Accretion discs supported by magnetic pressure in particular
are geometrically thick at all luminosities, with inflow times as
short as the observed few year timescales in extreme variability
events to date. Future time-resolved, multi-wavelength observations
can distinguish between inflow through a geometrically thick disc as proposed here, and
alternative scenarios of extreme reprocessing of a central source or
instability-driven limit cycles.
\end{abstract}




\section{Introduction}

The standard ``thin disc'' theory 
\citep{shaksun1973,novthorne} explains the high radiative efficiency,
luminosity, and spectral peak locations seen from black holes accreting at the Eddington rate. However, it is difficult to reconcile the theory with optical/UV observations of active galactic nuclei (AGN):

\begin{itemize}
\item The variability is often nearly simultaneous over a wide range
  of wavelength, requiring a coordination speed close to the speed of
  light \citep{claveletal1991,kroliketal1991}, inconsistent with
  viscous inflow through a thin disc.

\item Observed accretion discs are several times 
  larger than predicted, as measured either by quasar microlensing
  \citep[][]{morganetal2010} or continuum reverberation
  \citep[][]{mchardyetal2014}.

\item AGN accretion disc spectra are broader than expected
  \citep[][]{zhengetal1997,davisetal2007}, typically peak around $1000$\AA\,
  \citep[][]{shulletal2012}, and do not show the expected $T_{\rm eff} \sim (\dot{m}/M)^{1/4}$\, dependence
  \citep{davisetal2007,laordavis2014}, where $M$ is the black hole
  mass and $\dot{m} \equiv 0.1 \dot{M} c^2 / L_{\rm Edd}$ is the
  dimensionless accretion rate, equivalent to the Eddington ratio
  $L/L_{\rm Edd}$ for a radiative efficiency $\eta = 0.1$. 

\item Thin accretion discs should be subject to thermal
  \citep{shakurasunyaev1976} and inflow \citep{lightmaneardley1974}
  instabilities in the inner disc, and should be truncated at large
  radius by gravitational instability
  \citep{shlosmanbegelman1987,goodman2003}. 

\end{itemize}

AGN optical/UV variability has typical rms amplitudes of $\simeq
10-20\%$. Wide field surveys find that up to $\simeq 50\%$ of quasars undergo 
``extreme'' variability, where the optical luminosity changes by a factor
$\gtrsim 2$ \citep{rumbaughetal2018}. In ``changing look'' AGN,
the broad emission lines appear or disappear, causing
transitions between Types 1 and 2 (or 1.8/1.9). Known previously in a few nearby Seyferts
\citep{tohlineosterbrock1976,cohenetal1986,storchibergmannetal1995},
such events are now also found
from luminous quasars in wide-field surveys
\citep{macleodetal2016,ruanetal2016,yangetal2017,wangetal2018} increasing the
sample to $\approx 40$ total. The changes in broad line flux correspond to factor $\simeq 10$
changes in continuum luminosity, which occur on timescales of a few
years \citep[e.g.,][]{lamassaetal2015,gezarietal2017}. The timescale does not seem to
depend on the type of object or whether the luminosity is increasing or decreasing.

Variable obscuration is disfavored for changing
look AGN. The obscuring medium would need to have a large size and high speed to cover both the broad line and continuum emission
regions. In one case the mid-infrared flux decreases, so that obscuring material would have to block the torus \citep[][]{sternetal2018}, while in another there is no linear polarization signature \citep{hutsemekersetal2017} as seen in Type 2 AGN. In addition, in
the Seyfert galaxy Mrk 1018 a full optical to X-ray SED shows no sign
of variable obscuration, but is rather fully consistent with a drop in
intrinsic accretion power by a factor $\simeq 10$
\citep{mcelroyetal2016,husemannetal2016} over a few years. The
timescales seen in the small sample of changing look events are 
similar to those of normal \citep{kellyetal2009,macleod2010} and
extreme AGN variability, and the extreme objects do not stand out from
the overall population except for their large variability amplitudes
\citep{rumbaughetal2018}. Therefore, changing look objects may simply
be the high amplitude tail of normal AGN
variability and not the result of discrete events \citep[e.g.,
mergers or state transitions,][]{kimetal2018,nodadone2018,rossetal2018}. Apparently many or even
most AGN can undergo rapid, large amplitude, coordinated luminosity
variations, implying a much faster propagation timescale through the
accretion disc than predicted: a ``viscosity crisis" \citep{lawrence2018}. 

The crisis comes from the fact that in standard theory the disc is expected to be ``razor
thin," but that theory is clearly problematic for
AGN. Here we estimate inflow times for three alternative disc models
(\S\ref{sec:geom-thick-agn}) introduced in the literature in order to
explain one or more of these tensions between theory and
observation. Generically, discs in these scenarios are thicker,
leading to shorter inflow times that could help explain extreme AGN
variability (\S\ref{sec:inflow-times-extreme}). We show that accretion
discs supported vertically by strong magnetic fields (``magnetically
elevated discs") provide a particularly promising explanation, and
discuss how time-resolved photometry of changing look events can 
constrain their physical origin (\S\ref{sec:discussion}). 

\section{Geometrically thick AGN accretion discs}\label{sec:geom-thick-agn}

In the standard thin disc model \citep{shaksun1973,novthorne},
radiation pressure provides the vertical support against gravity at
small radius \citep{krolik1999},

\begin{equation}
R \lesssim 1000 \hspace{2pt} \alpha^{2/21} \left(\frac{L}{10^{46}
    \hspace{2pt} \rm erg \hspace{2pt} \rm s^{-1}}\right)^{2/21}
\left(\frac{\kappa}{\kappa_T}\right)^{20/21} \dot{m}^{2/3} r_g,
\end{equation}

\noindent where $\alpha \simeq 0.02$ \citep{hawleyetal2011} is the dimensionless viscosity parameter, 
$\kappa$ is the opacity scaled to the value for Thomson scattering,
and $r_g = GM/c^2$ is the gravitational radius. Both the predicted and measured
\citep{daietal2010,blackburneetal2014,mchardyetal2014} sizes of the
X-ray and optical emission regions for $\dot{m} \gtrsim 10^{-2}$ are
in the radiation-dominated regions of the disc. 

When radiation pressure dominates, the disc is thin, with a constant
height $H$ for radii $R$ far from the inner edge:

\begin{equation}
{\frac{H}{R}} = \frac{3}{2} \frac{\kappa}{\kappa_T} \frac{L}{L_{\rm Edd}} \frac{1}{\eta} \frac{\mathcal{R}_R}{\mathcal{R}_z} \frac{r_g}{R},
\end{equation}

\noindent where $\mathcal{R}_R$ and $\mathcal{R}_z$ are relativistic correction factors
\citep{novthorne,krolik1999} approaching unity at large $r/r_g$ and $\mathcal{R}_R = 0$
at $R/r_g = r_{\rm in}$ where $r_{\rm in}$ is the disc inner radius
assumed to be the marginally stable orbit. 

The inflow time is given as $t_{\rm inflow} \simeq R^2 / \nu$, where
$\nu$ is the disc viscosity. In the thin disc framework, this is 

\begin{equation}
\begin{aligned}\label{eq:inflowtime}
&t_{\rm inflow} \simeq \left(\Omega \alpha\right)^{-1} \left(H/R\right)^{-2},\\
&t_{\rm inflow} \simeq 500 \left(\frac{\alpha}{0.02}\right)^{-1}
\left(\frac{\kappa}{\kappa_T}\right)^{-2} \left(\frac{M}{10^8
    M_\odot}\right) \left(\frac{\dot{m}}{0.1}\right)^{-2}
\left(\frac{R}{50 r_g}\right)^{7/2} \hspace{2pt} \rm yr,
\end{aligned}
\end{equation}

\noindent where the second estimate is for the thin disc model in the
radiation pressure supported regime far from the inner edge, and we
fix the black hole mass $M = 10^8 M_{\odot}$ throughout.

This timescale is long at large radii and/or low luminosity, so that
changes in the outer optical emission region should be much slower and
separated in time from those in the inner UV and X-ray region. Neither
should operate on timescales of a few years as seen in changing look
AGN. Equally problematic, the inflow time should scale strongly with
luminosity, in conflict with observations where ``turn on" and ``turn
off" events occur equally rapidly \citep{lamassaetal2015,gezarietal2017}.

Proposed alternative accretion disc models remove
inconsistencies in standard theory and help reconcile it with optical/UV observations. The disc structure is generically thicker in these models, leading to shorter inflow times. Here we estimate inflow times of a few such scenarios which hold promise for explaining extreme AGN variability.

\subsection{Extra dissipation near the inner edge}

\citet{shaksun1973} and \citet{novthorne} assumed that the stress
vanishes at the disc inner edge. This is convenient for calculating
disc structure, but leads to singularities in the surface density and
scale height which disappear in a more careful treatment of the
boundary conditions \citep{abramowiczetal1988}. \citet{gammie1999} and
\citet{agolkrolik2000} further examined the effects of magnetic
torques, which do not vanish at the inner edge.

In \citet{agolkrolik2000}, the effect is parameterized as an extra radiative efficiency $\Delta \eta$ which contributes to $\mathcal{R}_R$:

\begin{equation}
\mathcal{R}_{R,\rm AK} = \frac{C_{\rm in} r_{\rm in}^{3/2} r_g^{1/2}}{C R^{1/2}} \Delta \eta + \mathcal{R}_R,
\end{equation}

\noindent where $C$ is another relativistic correction factor
\citep{agolkrolik2000} of order unity and $C_{\rm in}$ is evaluated at
$r_{\rm in}$. The disc scale height increases significantly when
$\mathcal{R}_{R, \rm AK} - \mathcal{R}_R \gtrsim 1$ or for $R/r_g < \left(\Delta \eta \hspace{2pt} C_{\rm
    in}/C \right)^2 r_{\rm
  in}^{3}$. For a non-spinning black hole, the disc
thickness significantly increases for $R_{\rm th} \lesssim 15$, $200 r_g$ for $\Delta \eta =
0.25$, $1$. The effects are strongly concentrated near $r_{\rm in}$
since the extra dissipation term scales as $H \sim R^{-1/2}$
instead of $H \sim R^0$ for a thin disc. 

\subsection{UV line opacity}

The thin disc solution includes free-free and electron scattering
opacity. The temperatures of inner AGN discs are similar to those of
massive stars, and so line opacity from heavy elements is expected to
play an important role. When including lines (ignored in the thin disc
model), the mean opacity at temperatures $T \sim 10^5$
K is a factor $\simeq 2-3$ higher than
$\kappa_T$ for solar metallicity \citep{jiangetal2015}. This can alter the
thermal stability of the disc \citep{jiangetal2016}, and clearly including
it will increase the scale height at least over the range of radii with
temperatures near this value.

As in massive stars, line driving could also produce powerful winds \citep{murraychiang1995,progaetal2000,proga2005}. \citet{laordavis2014} showed that if the
mass loss rate per unit area scaling derived in O stars holds for AGN,
the total mass loss rate exceeds the accretion rate inside the radius

\begin{equation}
R_{\rm eq} \sim 41.5 \left(\frac{M}{10^8 M_\odot}\right)^{0.24}
\dot{m}^{0.24} r_g.
\end{equation}

\noindent In their model, the disc truncates at $R_{\rm eq}$. At
smaller radii the disc becomes geometrically thick and 
optically thin, greatly decreasing the inflow time for $R 
\lesssim R_{\rm eq}$. 

A proper calculation of the scale height from this model would require
self-consistently solving for accretion and outflow, whereas currently
simulations including the frequency-dependent line opacity treat the thin
disc as a boundary condition and ignore its vertical structure
\citep[e.g.,][]{progaetal2000}. Radiation MHD simulations including line opacity are now possible \citep{jiangetal2017}, but use a gray opacity. Here we instead use a schematic picture of such a disc to estimate the inflow
time. We assume $H/R = 1$ for $R \le R_{\rm eq}$. The effects of
mass loss are highly concentrated near $R_{\rm eq}$
\citep{laordavis2014}, so we adopt a 
power law scaling of $H/R \sim R^{-\beta}$ and find $\beta$ by
matching onto the thin disc value of $H/R$ at $2 R_{\rm eq}$.

\subsection{Magnetically elevated discs}

The thin disc model includes gas and radiation pressure but ignores
magnetic fields. MHD simulations of small patches of accretion discs
(shearing box simulations) generically find
that magnetic pressure declines more slowly with height than radiation
or gas pressure, and so dominates in the upper atmosphere of the disc
\citep{millerstone2000}. Those simulations usually adopt an
initial condition with weak or no vertical magnetic field. If it can
be efficiently brought to the black hole, the magnetic flux available
in the inner parts of galaxies could instead be large. In that case,
the toroidal field amplified by the MRI \citep{mri} can become strong enough to support the disc vertically \citep{baistone2013,salvesenetal2016}. 

The vertical structure of such magnetically elevated discs remains uncertain.  In an early one-zone model,  \citet{begelmanpringle2007} assumed that the toroidal field grows
to a limiting field strength beyond which the MRI shuts off \citep{pessahpsaltis2005}: $v_A
\simeq \sqrt{2 c_s v_K}$, where $v_A$, $v_K$, and $c_s$ are the
Alfv\'{e}n, orbital, and sound speeds. The gas to magnetic pressure
ratio is $\beta \sim c_s / v_K$, which is $\sim H/R$ for a gas
pressure supported thin disc and so is small.

\citet{begelmanetal2015} took into account the competition between the
generation of toroidal field by MRI and shear --- assumed to occur at
all heights --- and its buoyant escape.  Their vertically stratified
models agree better with shearing-box simulations
\citep{baistone2013,salvesenetal2016} and yield a much larger
characteristic scale height, even if the Pessah--Psaltis
criterion is imposed: 

\begin{equation}\label{eq:1}
\frac{z_2}{R} \sim 0.75 \alpha^{-0.5} \dot{m}^{0.4},
\end{equation}

\noindent where we have used their height $z_2$ for the elevated
MRI-active layer. More recent studies of MRI in
the presence of a strong toroidal field, however, show that MRI never
completely stabilizes, even for very strong toroidal fields, although
it passes through a range of small values  \citep{dasetal2018}.
Moreover, shearing-box simulations, which lack toroidal field-line
curvature and should be more stable than global simulations with a
strong toroidal field \citep{blaesbalbus1994}, show no evidence of MRI
suppression in the nonlinear state.  It seems reasonable to guess that
magnetically elevated discs can thicken to $H/R \gtrsim 0.1$
\citep{begelmansilk2017}. In the following we will adopt the result of
equation \ref{eq:1}, as well as constant $H/R = 0.1$ for comparison. The MRI stress
also grows with magnetic flux, and in the magnetically elevated case we
assume $\alpha = 0.3$. However, $t_{\rm inflow}$ is independent of
$\alpha$ when the height is given by equation \ref{eq:1}.

\begin{figure}
	\includegraphics[width=\columnwidth]{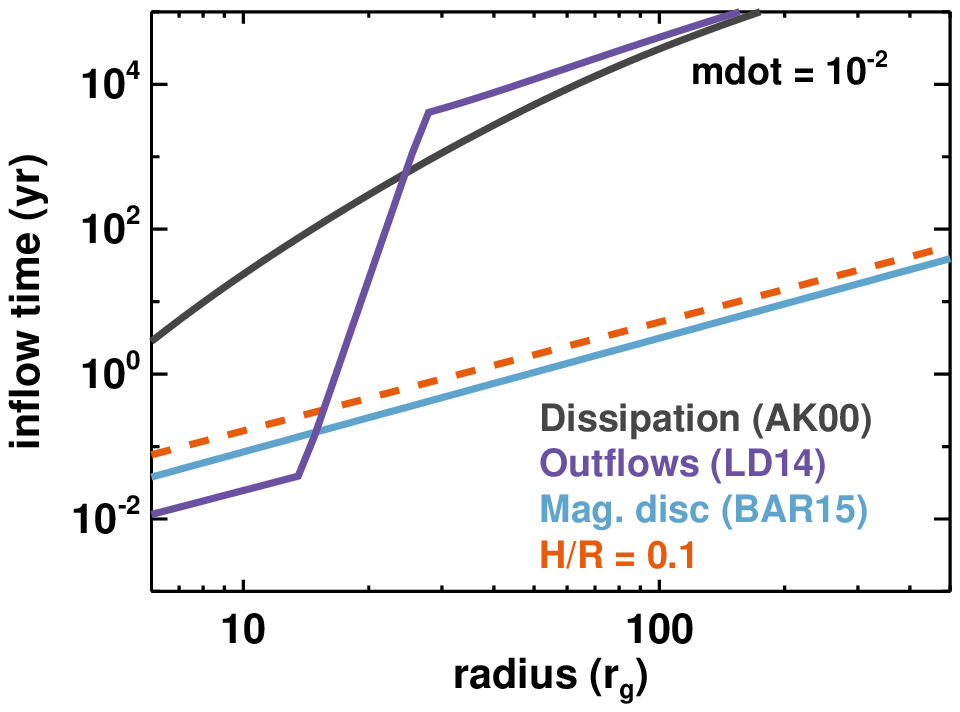}\\
    \includegraphics[width=\columnwidth]{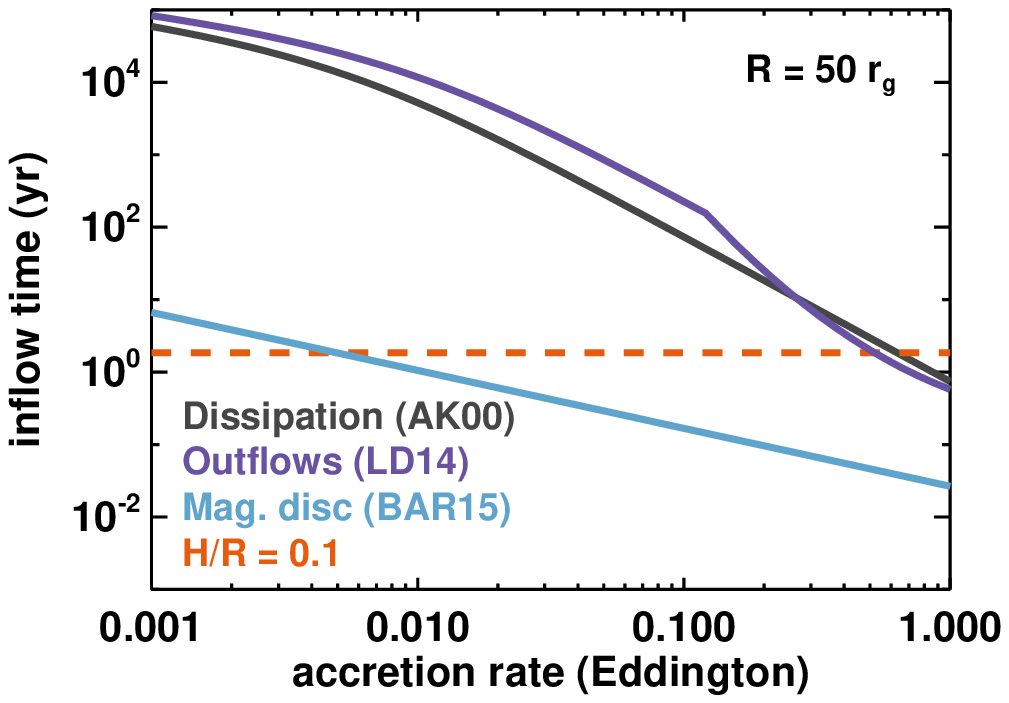}
    \caption{Estimated inflow times as a function of radius ($R/r_g$, top) and
      accretion rate ($\dot{m}$, bottom) for $M = 10^8 M_\odot$. A standard thin disc with extra
      dissipation at the ISCO (gray) becomes considerably thicker near
      the ISCO, but the inflow time at the optical emission region is
      mostly unaffected. UV line opacity could drive a strong wind and
      support a thick inner disc. The wind is expected to be strongest
      at small radius and high $\dot{m}$, and again may not alter the
      optical emission region. Magnetically dominated discs are
      generically geometrically thick, leading to much shorter inflow
      times $\lesssim 10$ years over a wide range of luminosity
      (blue). The \citet{begelmanetal2015} estimate we adopt gives
      similar results to a constant $H/R = 0.1$ (dashed orange).}
    \label{fig:tinflow}
\end{figure}

\subsection{Inflow times and extreme AGN variability}\label{sec:inflow-times-extreme}

Figure \ref{fig:tinflow} shows estimated inflow times (equation
\ref{eq:inflowtime}) for each of these scenarios as a function of
$R/r_g$ and $\dot{m}$ for $M = 10^8 M_\odot$ ($t_{\rm inflow} \propto
M$). For low accretion rates and large radii, the
thin disc inflow time is very long ($\gtrsim 10^3$ years) even when
including extra dissipation or possible effects of line-driven
outflows (dark gray and purple curves). In both scenarios the inflow
time becomes short for $R \lesssim 20 r_g$ or $\dot{m} \gtrsim 0.1$
where the added effects become important ($R < R_{\rm th}$ or $R_{\rm
  eq}$). For intrinsic disc emission at the measured size $R \simeq 50 r_g$
\citep{daietal2010,blackburneetal2014,mchardyetal2014}, changing look
AGN in relatively low-luminosity Seyfert galaxies 
($\dot{m} \sim 10^{-3}-10^{-2}$) are difficult to explain in these scenarios.

Magnetically elevated discs are geometrically thick
over the entire parameter space, leading to very short inflow times
(blue). When magnetic pressure supports the disc vertically against
gravity, its thermal content (luminosity, temperature) is decoupled
from the structure. The scale height and inflow time are not only
short, but also relatively insensitive to $\dot{m}$. This provides a
natural explanation for the few year timescales seen in changing look
AGN at both low and high luminosity. 

\section{Discussion}\label{sec:discussion}

\begin{figure}
	\includegraphics[width=\columnwidth]{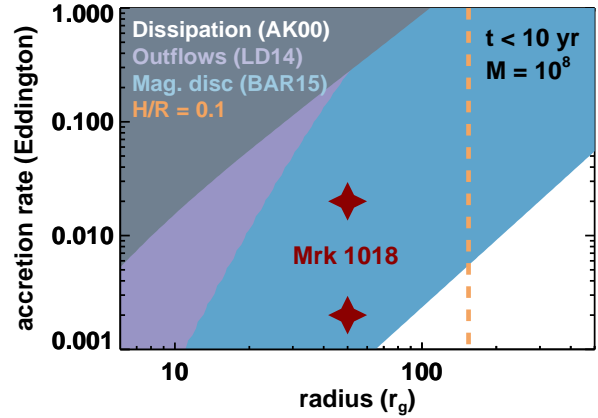}
    \caption{Allowed parameter space ($t_{\rm inflow} \le 10$ years,
      shaded regions) for changing look AGN as a function of radius
      and accretion rate assuming $M = 10^8 M_{\odot}$. The red stars correspond to the high and
      low states of the changing look Seyfert galaxy Mrk
      1018 \citep{mcelroyetal2016}. In either a standard thin disc or including extra
      dissipation at the ISCO, the inflow time is long except in the
      inner disc at high luminosity ($L/L_{\rm Edd} \gtrsim 0.1$). UV
      opacity could further inflate the inner disc, but may not be
      effective at the low luminosities of changing look Seyferts or
      in the outer optical emission regions. Magnetically dominated
      discs are expected to be geometrically thick over a wide range
      of radius and luminosity (blue, $H/R = 0.1$ dashed orange line for
      comparison) and provide a natural explanation 
      for rapid large amplitude variability.}
    \label{fig:tinflow_contour}
\end{figure}

Extreme AGN variability with factor $2-10$ optical/UV/X-ray luminosity changes on $\simeq 1-10$ year timescales places stringent constraints on accretion theory. In the standard picture of fluctuations propagating through the disc, the propagation timescale must be much shorter than expected: the disc must be geometrically thick. We have shown that three alternative AGN accretion disc models from the literature predict larger scale heights and shorter 
inflow times than in standard theory. Figure \ref{fig:tinflow_contour} summarizes the results,
showing regions in the $\dot{m}$ ($L/L_{\rm Edd}$) and $R/r_g$ parameter
space where our estimated $t_{\rm inflow} \le 10$ years in each
case, as implied \citep{pringle1981} by the $\lesssim 1$ yr timescale for the most rapid large amplitude variations seen so far \citep{gezarietal2017}. The red stars schematically show the changing look AGN Mrk 1018, assuming that the variations are sourced at the optical emission radius $R = 50 r_g$. Adding extra dissipation only leads to short inflow times at small radius and/or high luminosity. The result is similar for the
effects of UV opacity \citep{laordavis2014}. Magnetically elevated discs are a particularly
appealing option for reconciling extreme AGN variability with disc
theory. The disc is geometrically thick leading to short inflow times
with a weak dependence on luminosity, consistent with the observed few
year timescales in both Seyfert galaxies and quasars.

Our estimates are heuristic, either taken from past semi-analytic work
\citep{agolkrolik2000,begelmanetal2015} or estimated in toy models
based on that work \citep{laordavis2014}. In the case of UV opacity,
we have not calculated the vertical structure, and have ignored the
bulk increase in the total opacity coming from lines
\citep[e.g.,][]{jiangetal2015}. A calculation of the disc structure
and emergent spectrum are needed to determine whether a magnetically
elevated disc can produce the observed AGN 
SED, especially at high $\dot{m}$ where the absorption and effective
optical depths may be small. These improvements are left to future
work. 

Geometrically thick accretion discs help explain how large amplitude
AGN variability can occur on timescales $\simeq 1$ year, but the
physical origin of the events remain uncertain. Scenarios proposed for
changing look AGN include instability driven limit cycles
\citep{rossetal2018} or state transitions \citep{nodadone2018}, both
of which are seen in X-ray binaries. These scenarios invoke radiation
pressure effects, which we disfavor. Radiation pressure is strongest
at high luminosity and small radius, whereas extreme variability
prefers low luminosity \citep{rumbaughetal2018} and the optical
emission comes from large radius
\citep{morganetal2010,mchardyetal2014}. The short observed timescales
can be explained if the variability is driven entirely by reprocessed
UV/X-ray radiation from the inner disc
\citep{shappeeetal2014,lamassaetal2015,lawrence2018} and radiation
pressure and outflow will increase the irradiation by thickening the
inner disc \citep{agolkrolik2000}. However, this seems disfavored by
energetics in Mrk 1018 due to the low X-ray/UV luminosity
\citep{husemannetal2016}. In addition, in all of these scenarios the
variability timescale should strongly depend on luminosity
(eq. \ref{eq:inflowtime}), which has not been seen.

We propose instead that AGN accretion discs are supported by strong
toroidal magnetic fields. Magnetic pressure support leads to a
geometrically thick disc whose scale height is decoupled from its
thermal properties. This suppresses the thermal instability and leads
to a short inflow time at all observed luminosities. In this scenario,
the variability mechanism is the same for normal and extreme AGN, with
the latter the high amplitude tail of a continuous
distribution. \citet{rumbaughetal2018} find that extreme AGN make up
$\simeq 30-50\%$ of the population, and are only distinct in their
variability amplitudes (and slightly lower $\dot{m}$). The variability
mechanism is likely to either be mass accretion rate or thermal
fluctuations \citep{kellyetal2009,ruanetal2014,hungetal2016} which
then propagate through the disc to produce changes over a wide
spectral range. Magnetically elevated
accretion shows large density inhomogeneities \citep{salvesenetal2016}
and could support large temperature fluctuations. The same features
may produce the observed flat spectra and large sizes in the
optical/UV \citep{dexteragol2011,halletal2018}. For the observed
$\simeq 1$ year timescales, this scenario predicts $R \sim 10^{2} r_g$
for the origin of the variability (figure \ref{fig:tinflow_contour}), in the outer disc but within the
broad emission line region. The broad line luminosity and width should
therefore trace the continuum variations.

A sharp change in mass accretion rate should be seen first in the
optical and finally in the UV as it propagates inwards. In the case of
extreme reprocessing, the propagation should proceed with high 
energies leading, while instability-driven heating and cooling fronts
can travel in both directions. In this way time-resolved, multi-band photometry 
during extreme variability events can provide a direct probe of AGN
accretion physics.

\section*{Acknowledgements}

JD thanks S. W. Davis and the participants of the 2017 meeting ``Unveiling the Physics
Behind Extreme AGN Variability" for stimulating discussions. This work
was supported by a Sofja Kovalevskaja award from the Alexander von
Humboldt foundation. MB acknowledges support from NASA Astrophysics Theory Program grant  NNX17AK55G.




\bibliographystyle{mnras}








\bsp	
\label{lastpage}
\end{document}